\documentclass[twocolumn,superscriptaddress,floatfix,prl, aps]{revtex4-1}
\usepackage{graphicx,amsfonts,amssymb,amsmath,hyperref,hypcap,enumerate}
\usepackage{xcolor}
\usepackage{soul}
\definecolor{mygray}{gray}{0.6}
\usepackage{newtxtext,newtxmath} 
\usepackage{scalerel}

\newcommand{\sub}[1]{\scaleto{\text{#1}}{4pt}}

\begin{document}

\title{Voltage-Controlled Magnetic Reversal in Orbital Chern Insulators}
\author{Jihang Zhu}
\affiliation{Department of Physics, University of Texas at Austin, Austin TX 78712}
\author{Jung-Jung Su} 
\affiliation{Department of Electrophysics, National Chiao Tung University, Hsinchu 300, Taiwan}
\author{A. H. MacDonald}
\affiliation{Department of Physics, University of Texas at Austin, Austin TX 78712}

\begin{abstract}
Chern insulator ferromagnets are characterized by a quantized anomalous Hall effect (QAHE), 
and have so far been identified experimentally in magnetically-doped topological insulator (MTI) thin films and 
in bilayer graphene moir{\' e} superlattices.
We classify Chern insulator ferromagnets as either spin or orbital, depending on whether the orbital magnetization (OM)
results from spontaneous spin-polarization combined with 
spin-orbit interactions, as in the MTI case, 
or directly from spontaneous orbital currents,
as in the  moir{\' e} superlattice case.  We argue that in a given magnetic state, characterized for example by the 
sign of the anomalous Hall effect (AHE), the magnetization of an orbital Chern insulator
will often have opposite signs for weak $n$ and weak $p$ electrostatic or chemical doping.
This property enables pure electrical switching of a magnetic state in the presence of a fixed 
magnetic field.
\end{abstract}

\maketitle

{\em Introduction---}
A ferromagnet may be defined as an equilibrium state of matter in which time-reversal (TR) symmetry
is broken without lowering translational symmetries.  Ferromagnets generically have both 
non-zero spin magnetization and non-zero OM.  In almost all ferromagnets, the microscopic 
mechanism responsible for order is spontaneous spin-alignment 
driven by exchange interactions, which breaks spin-rotational invariance and leads to a non-zero 
spatially averaged spin moment density.  
Spin-orbit interactions then play a 
secondary role by inducing a small parasitic contribution to magnetization from orbital currents and a 
related non-zero (anomalous) Hall conductivity. 

This Letter is motivated by recent experiments\cite{XueQAHE, GoldhaberGordon2019BL, GoldhaberGordonWang2019TL, Young2019QAHE, Young2020QAHE_tMBG, Yankowitz2020AHE_tMBG}
that have established the QAHE in two quite different classes of two-dimensional ferromagnets.
The QAHE signals\cite{Thouless,Haldane} the formation of a ferromagnetic state, often referred to as a Chern insulator,
with occupied quasiparticle bands whose topological Chern numbers\cite{niu2010} sum to a non-zero value.
We find that when ferromagnetism mainly results from spontaneous 
orbital moments (not spin moments), as in the QAHE states 
recently discovered\cite{GoldhaberGordon2019BL, GoldhaberGordonWang2019TL, Young2019QAHE} in magic angle twisted bilayer 
graphene (MATBG), the magnetizations of weakly $n$-doped and weakly $p$-doped 
insulators can differ in sign in the same magnetic state characterized for example by a given sign of the anomalous Hall conductivity. This property could enable magnetic state reversal in the presence of a magnetic field to be 
achieved purely electrically.

The mechanism that allows the magnetizations of weakly $n$-doped and weakly $p$-doped Chern insulators to differ 
drastically is closely related to the quantum Hall effect itself.  Because of the presence 
of protected edge states, the OM $M$ of a 
Chern insulator 
changes\cite{LesHouches} with chemical 
potential $\mu$ even when $\mu$ is inside the bulk energy gap: 
\begin{equation}
\label{Eq:dMdmu}
\frac{dM}{d\mu} = \frac{dI}{d\mu} = \frac{ C e}{2 \pi \hbar},
\end{equation}
where $C$, the Chern index sum, is an integer equal to the Hall conductance in $e^2/h$ units.
Eq.~(\ref{Eq:dMdmu}) emphasizes that
the quantized Hall conductance can be 
understood\cite{LesHouches}
in terms of chiral edge states that are occupied to different chemical 
potentials along different portions of the sample boundary. 
It follows from Eq.~(\ref{Eq:dMdmu}) that the magnetization jumps by 
\begin{equation}
\label{Eq:DeltaM}
\Delta {M} = \frac{C e E_{\text{gap}}}{2 \pi \hbar}
\end{equation}
when the chemical potential jumps across the gap of a Chern insulator. 
Note that the jump in the magnetization
depends only on the value of the energy gap and on fundamental constants.   We show below that in orbital Chern insulator ferromagnets this jump can be sufficient to change the sign of magnetization simply by changing the sign of doping.

{\em Spin Chern Insulators---}  In MTI thin films, TR symmetry is broken by 
introducing local moments that order ferromagnetically. Spin-orbit coupling then leads to an AHE that is quantized, 
and to orbital ferromagnetism.  To compare the OM jump with the magnitude of the spin magnetization, 
we express it in 
units of Bohr magnetons $\mu_{\text{B}}=e\hbar/2m$ per surface unit cell: 
\begin{equation}
\label{Eq:DeltaMDimensionless}
\frac{\Delta M}{\mu_{\text{B}} /A_{\text{uc}}} = \frac{C m A_{\text{uc}} E_{\text{gap}}}{\pi \hbar^2}
\end{equation}
where $A_{\text{uc}}$ is the area of the surface unit cell.
In MTIs, spin magnetization in Bohr magnetons per surface unit cell is typically $\sim 1$, because the fraction of sites with 
magnetic atoms is $\sim  0.1$ and the number of magnetically doped layers is $\sim 10$.
Note that the spin magnetization 
does not depend on the position of the chemical potential within the gap.
We see from Eq.~(\ref{Eq:DeltaMDimensionless}) that the OM jump across the gap is small
compared with the spin magnetization 
since the surface state energy gap, although not known accurately, is certainly 
small compared  to the $\hbar^2/mA_{\text{uc}}$, which depends only on fundamental constants and the surface 
unit cell area and has a typical value in the $1-10$ eV range.  For MTIs, and other 
spin Chern insulators, the unusual jump in the magnetization across the insulator's gap is small in a relative sense and 
unlikely to have a qualitative influence on magnetic properties.

{\em Orbital Chern Insulators---}  
The Hall conductivity of a Chern insulator ferromagnet is quantized when the chemical potential lies in the gap or when
carriers introduced by chemical or electrostatic doping are localized. 
It is convenient to use the sign of the Hall conductivity to distinguish a magnetic state from its TR counterpart. We will refer to the state with positive quantized Hall conductivity $C e^2/h$ as the $+$ state 
and to the state with negative quantized Hall conductivity $-C e^2/h$ as the $-$ state.  
Although their variations with chemical potential are very distinct, as we emphasize below, both the Hall conductivity $\sigma_{\text{H}}^{\pm}(\mu)$ and OM $M^{\pm}(\mu)$
are orbital fingerprints of broken TR and at any doping level have opposite signs 
in TR partner states:
$
M^-(\mu) = - M^+(\mu), \  
\sigma_{\text{H}}^-(\mu) = - \sigma_{\text{H}}^+(\mu).
$

The TR symmetry breaking mechanism active in the orbital Chern insulators recently discovered in MATBG devices
has been actively discussed in recent work\cite{Bultinck_tBLG_AHE, Repellin_tBLG_AHE, Zhang_tBLG_AHE, Fengcheng_tBLG_magnon, Alavirad_tBLG_AHE, Liu_tBLG_AHE}.  It is almost certainly related to 
condensation in momentum space, a concept discussed some time ago by 
Heisenberg and London\cite{London}  and previously proposed\cite{JeilJungBilayer} as a 
possible symmetry breaking mechanism in metallic gated AB Bernal bilayer graphene.  
Momentum space condensation is driven by the property that
interaction energies in systems with long-range Coulomb interactions
can be lowered by occupying states that are more compactly distributed in momentum 
space than the occupied states of non-interacting bands.  
Just as exchange interactions in itinerant electron systems occur only between like spins,
exchange interactions between states with nearby momenta are stronger than those between 
states far apart in momentum space. In materials, like graphene, with low energy states located near two 
widely separated valley centers, momentum space condensation translates to spontaneous valley population polarization.
When combined with the intrinsically topological character\cite{vishwanath2019, Bernevig2019, Efetov2019}
of the valley-projected bands in these materials, valley polarization
yields an AHE that is quantized in insulating states.
The recently discovered graphene multilayer QAHE states\cite{GoldhaberGordon2019BL, GoldhaberGordonWang2019TL,Young2019QAHE} 
provide, as far as we are aware, the only demonstrated example of this mechanism at work.
In order to estimate the OM of these states we apply the convenient envelope function
description\cite{BistritzerPNAS}, in which the moir\'e superlattices is described by a valley-projected periodic 
Hamiltonian that accounts for position-dependent stacking.
We focus below on the case of twisted bilayer graphene (TBG) sandwiched by aligned
hexagonal Boron Nitride (hBN) layers.

{\em OM of MATBG on hBN---} 
The contribution to OM from a single band of 2D Bloch electrons
is\cite{niu2010, Vanderbilt2005,Vanderbilt2006,Bianco_OM_2013,Bianco_OM_2016}
\begin{eqnarray}\label{Mz_eqn}
    M_n(\mu) &=& \int \frac{d^2\pmb{k}}{(2\pi)^2} \mathcal{M}_n(\pmb{k},\mu) f(\mu-\varepsilon_n(\pmb{k})) \\
    \mathcal{M}_n(\pmb{k},\mu) &=& \frac{e}{\hbar} \text{Im} \sum\limits_{n^\prime \neq n} \frac{\langle n| \partial_xH| n^\prime \rangle \langle n^\prime |\partial_yH| n\rangle}{(\varepsilon_n-\varepsilon_{n^\prime})^2} \, (\varepsilon_n + \varepsilon_{n^\prime}-2\mu) \nonumber
\end{eqnarray}
where $n$ is a band index, $\mu$ is the chemical potential, $f(\mu-\varepsilon_n(\pmb{k}))$ is Fermi-Dirac distribution, 
$\partial_jH = \partial H/\partial k_j$ is the velocity operator and $|n\rangle$ is a Bloch state with implicit wave-vector 
dependence. We separate the OM in Eq.~(\ref{Mz_eqn}) into two parts by defining 
\begin{eqnarray}\label{M1M2}
M_n^1(\mu) &=& \frac{e}{\hbar} \text{Im} \sum\limits_{n^\prime \neq n} \int \frac{d^2\pmb{k}f_n}{(2\pi)^2} 
\frac{\langle n|\partial_xH|n^\prime \rangle \langle n^\prime |\partial_yH| n\rangle}{(\varepsilon_n-\varepsilon_{n^\prime})^2} (\varepsilon_n + \varepsilon_{n^\prime})  \nonumber\\
M_n^2(\mu) &=& \frac{e}{\hbar} \text{Im} \sum\limits_{n^\prime \neq n} \int \frac{d^2\pmb{k}f_n}{(2\pi)^2} \frac{\langle n|\partial_xH| n^\prime \rangle \langle n^\prime |\partial_yH| n\rangle}{(\varepsilon_n-\varepsilon_{n^\prime})^2} (-2\mu)
\end{eqnarray}
where $f_n$ is short for $f(\mu-\varepsilon_n(\pmb{k}))$.
When band $n$ is full, $M_n^1(\mu)$ is independent of $\mu$, whereas $M_n^2(\mu)$ includes the edge state contribution and is proportional to $\mu$ with proportionality constant $C_n e/2\pi \hbar$, where $C_n$ is the Chern number of band $n$.

We now apply these expressions to TBG encapsulated between hBN layers whose 
influence on the low-energy graphene Hamiltonian is
captured\cite{Wallbank2013, Koshino2014, JeilJung2015, JeilJung2017, JeilJung2014, Miller2012,Shi2020} in part
by a mass term representing the spatially averaged difference 
between carbon $\pi$-orbital energies on different honeycomb sublattices. 
The valley-projected TBG Hamiltonian is
$\mathcal{H}(\pmb{r}) =  h_0^{(1)} + h_0^{(2)}  + T(\pmb{r}) + h.c.$,  
where $h_0^{(l)}(\pmb{r}) = -i\partial_x \sigma_x - i\partial_y \sigma_y + m_l\sigma_z$ is the massive Dirac Hamiltonian of layer $l$, $\sigma$ acts on the
sublattice degrees of freedom, 
and $T(\pmb{r})$ is the periodic interlayer tunneling Hamiltonian\cite{BistritzerPNAS}. 
The conclusions we reach below rest in part on a particle-hole symmetry 
property of this Hamiltonian, discussed at greater length in supplementary material (SM) S1.
\begin{align}
    \label{symmetry1_r} &\tau_z \sigma_x \mathcal{H}(x,y) \sigma_x \tau_z = -\mathcal{H}(-x+d,y) \\ 
    \label{symmetry2_r} &\tau_x \mathcal{H}(x, y) \tau_x = \mathcal{H}^*(-x+d, y)
\end{align}
In Eqs.~(\ref{symmetry1_r},\ref{symmetry2_r}), 
$\tau$ acts on the layer degrees of freedom and 
$d=a_{\text{\tiny{M}}}/\sqrt{3}$ (modulo $\sqrt{3} a_{\text{\tiny{M}}}$) 
where $a_{\text{\tiny{M}}}$ is moir\'e lattice constant.
Symmetry (\ref{symmetry1_r}) states that up to a translation
and a change in the sign of the interlayer tunneling term,
sublattice exchange combined with reflection by the $y$-axis 
simply changes the sign of the Hamiltonian.
Eq.~(\ref{symmetry1_r}) becomes exact in the limit of small twist angles and is accurate in MATBG.
Eq.~(\ref{symmetry2_r}) is satisfied only when the masses of two graphene layers are 
identical. In momentum space, the Hamiltonian satisfies
\begin{align}
    \label{symmetry1} &\tau_z \sigma_x H(k_x, k_y) \sigma_x \tau_z = - H(-k_x, k_y) \\ 
    \label{symmetry2} &\tau_x H(k_x, k_y) \tau_x = H^*(k_x, -k_y) 
\end{align}
Given Eq.~(\ref{symmetry1}) it can be 
shown, as detailed in SM S1, that the contribution to OM
from a valley vanishes when $\mu$ lies in the middle of the gap between the 
conduction and valence bands of that valley.

\begin{figure}
  \includegraphics[width=1\columnwidth]{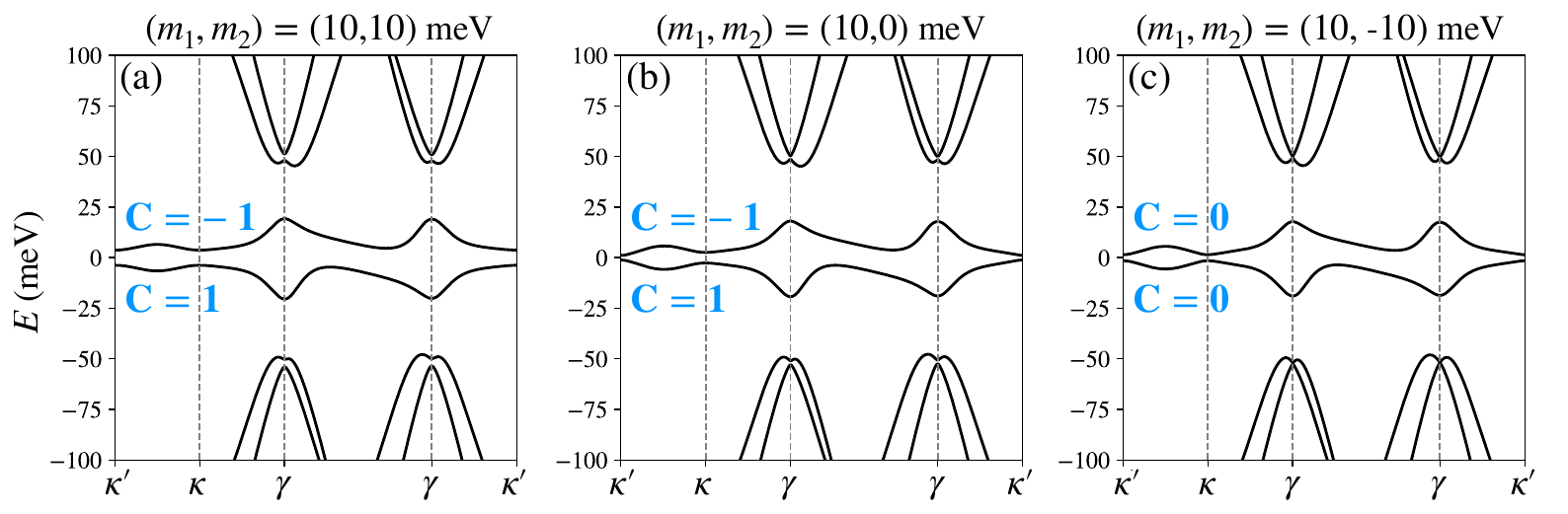}
  \vspace{-20pt}
  \caption{\small{
  $1.1^\circ$-TBG moir\'e band structures in valley $K$ for 
  three hBN-induced mass choices. (a) $m_1=m_2=10$ meV 
  produces a band gap $E_\text{g}^0 \sim 7.5$ meV at charge neutrality. 
  The flat bands are non-trivial with Chern numbers $C=\pm 1$. 
  (b) $m_1=10, m_2=0$ meV produces a band gap $E_\text{g}^0 \sim 2.3$ meV. 
  The flat bands are non-trivial with Chern numbers $C=\pm 1$. 
  (c) $m_1=-m_2=-10$ meV produces a band gap $E_\text{g}^0 \sim 3$ meV with
  topologically trivial flat bands.  The moir\'e bands
  were calculated using a low-energy continuum model\cite{BistritzerPNAS} with interlayer tunneling strength 
  $w^{\sub{AB}}=110$ meV and $w^{\sub{AA}}/w^{\sub{AB}}=0.85$ to account for corrugation and strain.}
  } 
  \label{band_structures}
\end{figure}

For graphene on hBN $m$ has been estimated using DFT\cite{JeilJung2015, JeilJung2017, JeilJung2014} 
to be $\sim 3.6$ meV for perfect alignment, but can be substantially enhanced by interaction 
effects absent in DFT and decreases with relative twist angle.  
Experimental $m$ values for nearly aligned graphene on hBN 
are $\sim$ 10$-$15 meV\cite{GhBN_gap_2018, GhBN_gap_2013, GhBN_gap_2019}.
Figure~\ref{band_structures} illustrates the $K$-valley low-energy moir\'e bands and  
Chern numbers of $1.1^\circ$-TBG for different mass choices.
The choice $m_1=m_2$ (Fig.\ref{band_structures}(a)) corresponds to the case in which both graphene layers are aligned
and have equivalent stacking orientation relative to their adjacent
hBN layers, while $m_1=-m_2$ (Fig.\ref{band_structures}(c)) corresponds to the case in which two  
graphene layers have opposite relative stacking orientations. 
$m_l=0$ (Fig.\ref{band_structures}(b)) corresponds to layer $l$ having a large 
misalignment relative to hBN so that strain enhancement is absent.
We find that gaps ($E_\text{g}^0$) appear at charge neutrality, 
that the bands are relatively flat for twists near the magic angle, and that they have non-zero 
Chern numbers when both layers have the same alignment or only one layer is aligned.
The case of opposite masses produces trivial bands (Fig.~\ref{band_structures}(c)). 
In all three cases sublattice-symmetry breaking gaps the 
Dirac points at the moir\'e Brillouin zone (MBZ) corners that otherwise link the 
conduction and valence bands.

\begin{figure}
  \includegraphics[width=1\columnwidth]{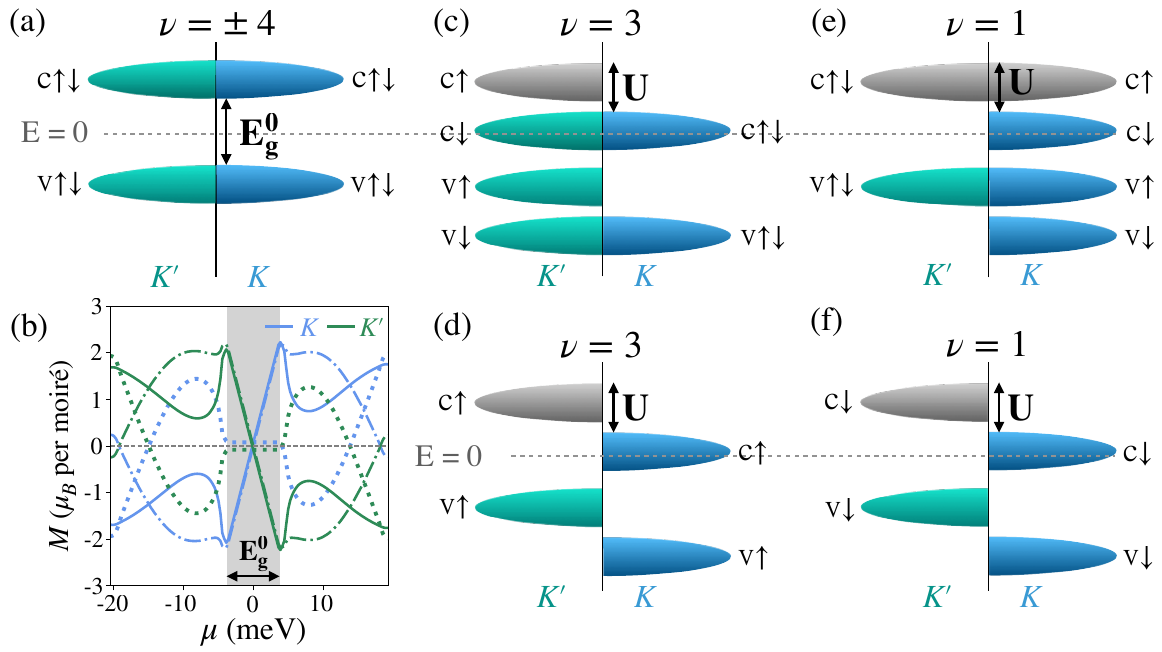}
  \vspace{-20pt}
  \caption{\small 
  (a) Schematic moir\'e flat bands with SU(4) spin/valley symmetry maintained. Each band has a four-fold degeneracy. $E_g^0$ is the single-particle band gap. "c" and "v" are shorts for "conduction" and "valence" bands.
  (b) Magnetization contributions (solid line) from valleys $K$ (blue) and $K'$ (green) as a function of $\mu$.
  The dotted and dash-dotted lines are respectively the $M^1$ and $M^2$ contributions defined in Eq.~(\ref{M1M2}). 
  The single-particle gap is shaded in gray.  
  (c-f) Schematic moir\'e flat bands in TR broken symmetry states at $\nu=3$ and $\nu=1$ 
  in which different flavors are rigidly shifted in energy by a momentum- and flavor-independent exchange energy $U$ if the flat conduction band is filled in that flavor.
  }
  \label{fig_valley_pola}
\end{figure}

{\em SU(4) symmetric mean-field model---}
Figure ~\ref{fig_valley_pola}(b) plots the single-flavor magnetization contributions (solid line)
from valleys $K$ and $K'$ at twist angle $1.1^\circ$ as a function of $\mu$ measured relative to the mid-point between its 
shifted conduction and valence bands. 
As explained previously the magnetization contribution from each valley vanishes at mid-gap and varies linearly within the gap.  Because valleys $K$ and $K^\prime$ are time-reversed counterparts, their magnetization contributions are always opposite in sign.
The dotted and dash-dotted lines in Fig.~\ref{fig_valley_pola}(b) separate the $M^1$ and $M^2$ contributions defined in Eq.~(\ref{M1M2}). 
The range of $\mu$ plotted in Fig.~\ref{fig_valley_pola}(b) covers from the flat valence band bottom to the flat conduction band top\cite{footnote1}.

Because of the four-fold spin/valley degeneracy of the moir\'e flat bands (Fig.~\ref{fig_valley_pola}(a)), gaps can appear 
only at moir\'e filling factors $\nu$ that are multiples of four when interactions are neglected.  
To account for the Chern insulator gaps at odd integer values of $\nu$, 
we use a simplified but still qualitatively reliable\cite{Ming_tBLG} mean-field model 
in which exchange interactions shift all the band energies of a given flavor en masse -- down when the flat conduction band is 
occupied and up when the flat valence band is emptied (Fig.~\ref{fig_valley_pola}(c-f)).
The band energy shift $U$ must exceed the band width $w$ in order for the gapped state to be self-consistent;
this Stoner criterion is easily satisfied near magic angle orientations because $w$ is extremely small.
Schematic ordered state bands for $\nu=3$ and $\nu=1$ are plotted in Fig.~\ref{fig_valley_pola}(c) and (e).
For three electrons per moir\'e period ($\nu=3$), the density at which
the QAHE has been most often observed to date, all the majority $\downarrow$ spin's flat bands are occupied and the 
magnetization contributions from its two valleys cancel. We can therefore 
consider only the minority $\uparrow$ spin bands shown in Fig.~\ref{fig_valley_pola}(d). 
Similarly, for one electron per moir\'e period ($\nu=1$), we can consider only the majority $\downarrow$ spin bands illustrated in Fig.~\ref{fig_valley_pola}(f).

Although the magnetization of an orbital Chern insulator can in principle reverse its sign at any filling factor, depending on the details 
of the Chern band, abrupt reversals {\it vs.} gate voltage occur only at integer $\nu$, where the magnetization has a large jump. 
We, therefore, focus on the magnetization jumps at $\nu=3$ and $\nu=1$; a similar analysis applies for $\nu=-3$ and $\nu=-1$.
The total OM is calculated by summing Eq.~(\ref{Mz_eqn}) over spin/valley flavors and bands:
\begin{equation}\label{Eq_M_tot}
    M(\mu) = \sum\limits_{m,f} \big(M^1_{mf} n_{mf} + \mu \frac{C_{mf} n_{mf}}{2\pi} \big) + U \sum\limits_{m, f\in f_{\text{shift}}} \frac{C_{mf}n_{mf}}{2\pi}
\end{equation}
where $m$ is a band index in the valley- and spin-projected continuum model, 
$f$ is a flavor index, $f_{\text{shift}}$ is the set of flavors that have had their 
energies shifted by $-U$, $n_{mf}$ is the band occupation, and $C_{mf}$ is the band Chern number. $M^1_{mf}$ is evaluated with the zero of energy located at the middle of the single-particle gap $E_g^0$ as in Fig.~\ref{fig_valley_pola}(a). 
In the last term in Eq.(\ref{Eq_M_tot}), we have used that the magnetization contribution of an occupied band 
changes by $-C_{mf} \delta E/2\pi$ when the band energy is rigidly shifted by $\delta E$.

Since the magnetizations and Chern numbers of time-reversed bands 
cancel, {\it i.e.} $M^1_{mK}$ $=$ $-M^1_{mK'}$ and $C_{mK}=-C_{mK'}$, it follows that at both $\nu=3$ and $\nu=1$
\begin{equation}\label{Eq_Mmu_part1}
    \sum\limits_{m,f} \big( M^1_{mf}n_{mf} + \mu \frac{C_{mf}n_{mf}}{2\pi} \big) = M^1_{\text{c}K} +  \frac{\mu C_{\text{c}K}}{2\pi}.
\end{equation}
The extra magnetization contribution from occupied bands that suffer an exchange energy shift $U$ is
\begin{equation}\label{Eq_shift_term}
    U \sum\limits_{m, f\in f_{\text{shift}}} \frac{C_{mf}n_{mf}}{2\pi}
    = \frac{U(C_{\text{c}K} + C_{\text{v}K} + C_{\text{v}'K})}{2\pi}
\end{equation}
where $C_{\text{c}K}$/$C_{\text{v}K}$ is the Chern number of the flat conduction/valence band in valley $K$ and $C_{\text{v}'K}$ is the total Chern number summed over all remote valence bands. 

\begin{figure*}
  \includegraphics[width=\linewidth]{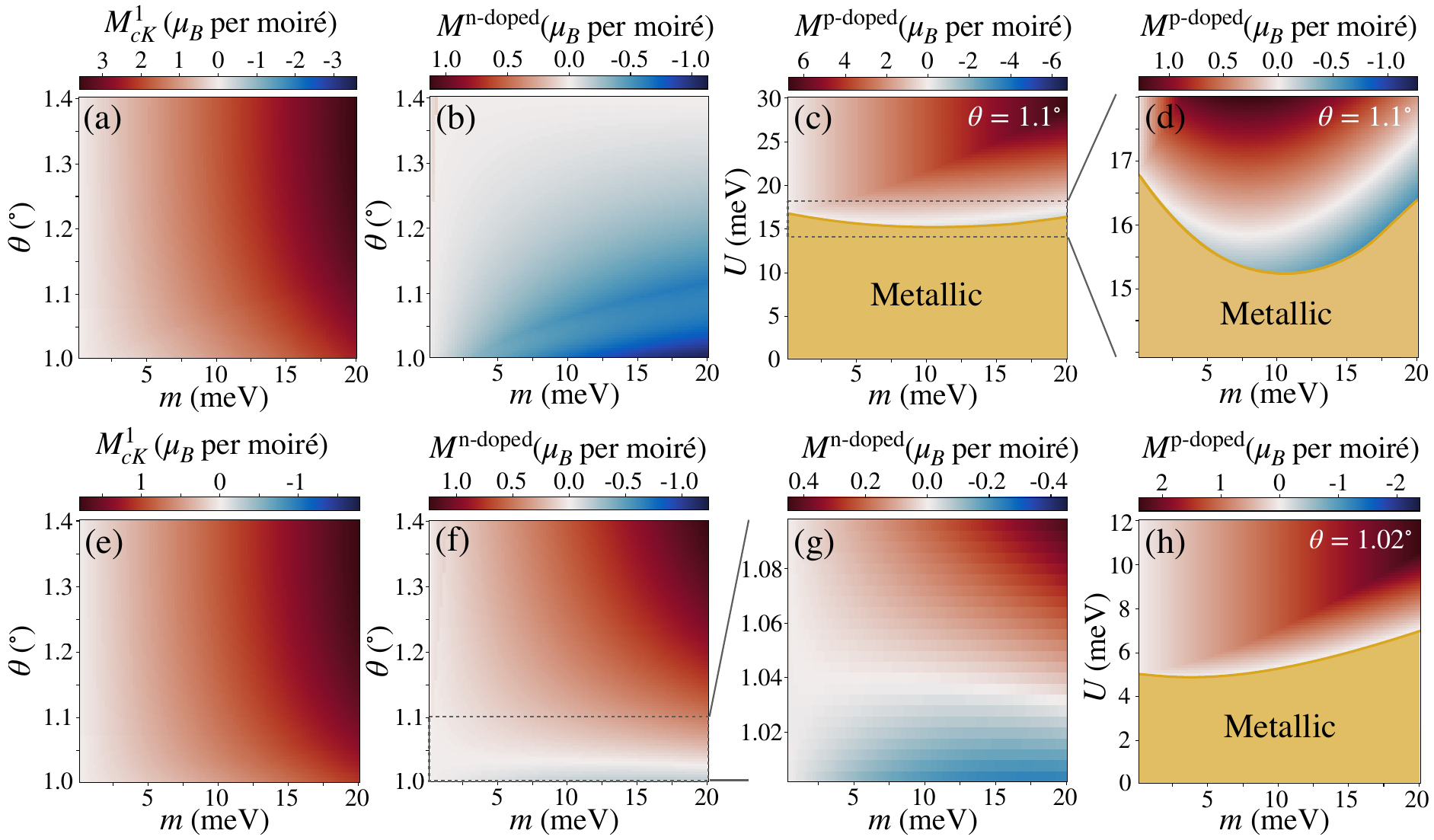}
  \vspace{-20pt}
  \caption{\label{fig3_M_2d} \small 
  Plots of $M^1_{\text{c}K}$, $M^{\text{n-doped}}$ and $M^{\text{p-doped}}$ defined in Eq.~(\ref{Eq_Mmu_part1}), (\ref{Eq_Mndoped}) and (\ref{Eq_Mpdoped}).
  (a-d) With $m_1=m_2=m$. (e-g) With $m_1 = m$ and $m_2=0$. (a) $M^1_{\text{c}K}$ increases as both twist angle $\theta$ and $m$. (b) $M^{\text{n-doped}}(\theta, m)$ is negative in most parts of the parameter range shown in the figure. (c) $M^{\text{p-doped}}(U,m)$ for a typical twist angle $\theta=1.1^\circ$. (d) Zoom-in of the dashed rectangle in (c). $M^{\text{p-doped}}$ is positive as long as $\Delta_g \gtrsim 1$ meV, which is easily achievable in MATBG. (e) Similar to (a), $M^1_{\text{c}K}$ increases as both $\theta$ and $m$. (f) $M^{\text{n-doped}}(\theta, m)$. (g) Zoom-in of the dashed rectangle in (f). $M^{\text{n-doped}}$ is only negative for $\theta \lesssim 1.04^\circ$. (h) $M^{\text{p-doped}}(U,m)$ for $\theta=1.02^\circ$. Similar to (d), $M^{\text{p-doped}}$ is positive for a tiny gap. In (c,d,h), the parameter region where $U<w$ is identified to be metallic.
  } 
\end{figure*}

In our simplified SU(4) symmetric model, $M^1_{\text{c}K}$ and the Chern numbers are purely single-particle properties. 
For the range of parameters ($\theta$, $m$) plotted in Fig.~\ref{fig3_M_2d}, $C_{\text{c}K}$=$-C_{\text{v}K}$=$-1$ and $C_{\text{v}'K}=0$. 
It follows that for both $\nu=1$ and $\nu=3$, the magnetization ($M^{\text{n-doped}}$) when $\mu$ is at the bottom of unoccupied band(s) is 
\begin{equation}\label{Eq_Mndoped}
    M^{\text{n-doped}} 
    = M(\mu=\frac{E_g}{2}) 
    = M^1_{\text{c}K} - \frac{E_g}{4\pi}
\end{equation}
and the magnetization ($M^{\text{p-doped}}$) when $\mu$ is at the top of occupied bands is
\begin{equation}\label{Eq_Mpdoped}
    M^{\text{p-doped}} 
    = M(\mu=\frac{E_g}{2}-\Delta_g) 
    = M^1_{\text{c}K} - \frac{E_g}{4\pi} + \frac{\Delta_g}{2\pi}
\end{equation}
where $\Delta_g = \text{min}\{U-w, E_g\}$ is the correlated gap at $\nu=1,3$. The magnetization sign reverses across the gap if 
\begin{equation}\label{Eq_require}
    M^{\text{n-doped}} < 0 \text{  and  }
    M^{\text{p-doped}} > 0
\end{equation}
In Fig.\ref{fig3_M_2d}(a) we show that $M^{1}_{\text{c}K}$ increases as a function of both twist angle $\theta$ and mass $m=m_1=m_2$.  
Figure \ref{fig3_M_2d}(b), which plots $M^{\text{n-doped}}(\theta, m)$, reveals that the first condition in Eq.(\ref{Eq_require}) 
is always satisfied near the magic twist angle.  Figure \ref{fig3_M_2d}(c) plots $M^{\text{p-doped}}$ for a typical twist angle $\theta=1.1^\circ$
{\it vs.} $U$ and mass $m$.  Insulating states occur only when $U>w$.  We find that $M^{\text{p-doped}}$ is almost always positive, satisfying the second condition in Eq.(\ref{Eq_require}), although there is a small no-reversal region in which the Chern insulator gap $\Delta_g=U-w \sim 1$ meV that is highlighted in Fig.~\ref{fig3_M_2d}(d).
Similar results for $m_1=m,m_2=0$ models are provided in Fig.~\ref{fig3_M_2d}(e-h). 
In this case, $M^{\text{n-doped}}$ is negative for $\theta \lesssim 1.04^\circ$, as illustrated in Fig.\ref{fig3_M_2d}(g).

{\em Discussion---}
Chern insulators are 2D electron systems with charge gaps that exhibit QAHE, and have now been realized experimentally by two distinct mechanisms.
In MTI\cite{XueQAHE, QAHE_exp_Checkelsky2014, QAHE_exp_Kou2014, QAHE_exp_Bestwick2015, QAHE_exp_Chang2015, QAHE_exp_Mogi2015, QAHE_exp_JueJiang2019}, the QAHE is driven by the exchange interactions between spin-local-moments that order ferromagnetically  
and two Dirac-cones localized on opposite surfaces of a topological insulator thin film.  
In bilayer graphene, on the other hand, the 
QAHE is driven by broken sublattice symmetry, which gaps Dirac cones and
induces Berry curvatures of opposite signs near TR-partner valleys,
combined with TR symmetry breaking via condensation of electrons into one of the two valleys.
Both experimentally established QAHE mechanisms differ from the one identified in the original theoretical work
of Haldane\cite{Haldane} in which the QAHE is driven by broken TR symmetry that leads to Berry curvatures 
of the same sign near opposite valleys.

In TBG sublattice polarization is theoretically expected to occur spontaneously,
but can be enforced by alignment with hBN.  Spontaneous valley polarization and
spin polarization are then energetically preferred when the moir\'e bands are narrowed by tuning the orientation
close to the magic angle.  Because of the absence of substantial spin-orbit coupling in graphene, the orbital 
valley order has Ising character and is therefore essential to achieve a finite transition temperature, and is dominantly 
responsible for the magnetization and solely responsible for the most accessible observable -- the 
QAHE.  We have shown in this Letter that the dominance of orbital magnetism change 
the considerations\cite{Dieny_PMA}
that normally limit our ability to control magnetic states electrically.  
The most extreme example of the strong electrical effects that are possible in orbital Chern insulators 
is a consequence of the jump in magnetization between weak $n$-doping and weak $p$-doping produced by edge states.
Changing the sign of magnetization of a state with a given sign of valley polarization and QAHE,
changes the thermodynamically preferred state in a weak magnetic field purely electrically.  
This property could be of technological value if other examples of orbital 
Chern insulators that have higher transition temperatures are discovered in the future.  When the sign of the magnetization 
is independent of carrier density, the St\v{r}eda\cite{Streda1982} formula implies that magnetic-switching between
quantum anomalous Hall states will yield stronger transport signals for either $n$- or $p$-doping, depending on the 
relative sign of magnetization and  
Hall conductivity. This behavior is common in current experiments\cite{GoldhaberGordonWang2019TL,Young2019QAHE, Efetov2019, CoryDean2019_WSe2, Efetov2019_CIandSC}.   
As illustrated in SM S3 QAHE sign switching that is equally robust for $n$- and $p$-doping signals the 
magnetization sign switch that we expect to be common in large gap orbital Chern insulators.

In our simplified mean-field theory the magnetizations at weak $n$- and $p$-doping are 
identical at $\nu=3$ and $\nu=1$ since Eqs.(\ref{Eq_M_tot}-\ref{Eq_Mpdoped}) apply to both cases.
This property is a consequence not only of the simplified mean-field theory but also of our neglect of correlations,
which are likely to play an important role in determining whether or not Chern insulator states appear.  
Since the flat-band system has more phase space for correlations closer to charge neutrality, we 
anticipate that Chern insulator states will be more common at $\nu=\pm 3$, than at 
$\nu=\pm 1$.

{\it Note added:} While this manuscript was under review, the magnetization sign reversal it predicts
was observed in twisted monolayer on bilayer graphene and in twisted bilayer graphene\cite{Polshyn_tMBG_2020}.

{\em Acknowledgements}---
JZ was supported by the National Science Foundation through the Center for Dynamics and Control of Materials: an NSF MRSEC under Cooperative Agreement No. DMR-1720595. 
AHM was supported by DOE BES grant FG02-02ER45958.  The authors acknowledge resources provided by the Texas Advanced Computing Center (TACC) at The University of Texas at Austin that have contributed to the research results reported in this paper.

\bibliographystyle{apsrev4-1}
\bibliography{main}

\begin{thebibliography}{59}
\bibitem [{foo()}]{footnote1}%
  \BibitemOpen
  \href@noop {} {\ }\bibinfo {note} {Note that $M^2$ (dash-dotted line) should
  vanish when the flat bands are completely filled or completely empty because
  the total Chern number of all occupied bands is zero in both cases. The
  numerical error in Fig.~\ref{fig_valley_pola}(b) is difficult to be
  completely eliminated because small inaccuracies in the Berry curvature
  integrals are magnified when multiplied by a finite chemical potential
  $\mu$.}\BibitemShut {Stop}%
\end{thebibliography}

\clearpage
\onecolumngrid
\section{Supplemental Material:\\ The Curious Magnetic Properties of Orbital Chern Insulators}

\section{S1. Symmetries of TBG}
Below we prove that the moir\'e Hamiltonian $\mathcal{H}(\pmb{r})$ of TBG on hBN substrates has 
the following symmetry properties:
\begin{align}
    &\tau_z \sigma_x \mathcal{H}(x,y) \sigma_x \tau_z = -\mathcal{H}(-x+d,y) \ \ \ \ \ \ \ \ \ \label{eq_s1}  \\
    &\tau_x \mathcal{H}(x, y) \tau_x = \mathcal{H}^*(-x+d, y)  \ \ \ \ \ \ \ \ \ \ \ \ \ \ \ \ \ \  \label{eq_s2}.
\end{align}
The first of these properties applies only in the limit of small twist angles and the second only if the 
Dirac masses $m_1$ and $m_2$ in the two layers are identical.  
The Pauli matrices $\sigma$ and $\tau$ act on sublattice and layer degrees of freedom respectively.

The moir\'e Hamiltonian
\begin{equation}
\mathcal{H}(\pmb{r}) = 
    \begin{pmatrix}
        h_0^{(1)}(\pmb{r}) & T(\pmb{r}) \\
        T^\dagger(\pmb{r}) & h_0^{(2)}(\pmb{r})
    \end{pmatrix}
\end{equation}
$T(\pmb{r})$ is the interlayer tunneling matrices and superscripts $(1)$ and $(2)$ denote graphene layers. $h_0(\pmb{r})$ is the 
massive Dirac Hamiltonian
\begin{equation}
    h_0(x,y) = -i\partial_x \sigma_x - i\partial_y \sigma_y + m\sigma_z
\end{equation}
and we have
\begin{equation}\label{tuneling_sigmax_h0}
    \sigma_x h_0(x,y) \sigma_x = -h_0(-x,y)
\end{equation}
For the interlayer tunneling term,
\begin{equation}\label{eqn_Tr}
    T(\pmb{r}) = w_0 \sum\limits_{j=1}^3 e^{-i\pmb{q}_j \cdot \pmb{r}} T_j
\end{equation}
where $\pmb{q}_1 = (0,-1), \pmb{q}_2 = (\sqrt{3}/2, 1/2), \pmb{q}_3 = (-\sqrt{3}/2, 1/2)$ (in units
of $4\pi/3a_{\text{\tiny{M}}}$) are three momentum boosts. $a_{\text{\tiny{M}}}$ is the moir\'e lattice constant.  $w_0=110$ meV is the tunneling strength, and
\begin{equation}
T_1=
    \begin{pmatrix}
    1 & 1 \\
    1 & 1
    \end{pmatrix},
\ \ \ \ \ T_2 =
\begin{pmatrix}
    e^{-i\phi} & 1 \\
    e^{i\phi} & e^{-i\phi}
    \end{pmatrix},
\ \ \ \ \ T_3 =
\begin{pmatrix}
    e^{i\phi} & 1 \\
    e^{-i\phi} & e^{i\phi}
    \end{pmatrix}
\end{equation}
with $\phi=2\pi/3$. 
Substituting $\pmb{q}_j$ and $T_j$ in Eq.~(\ref{eqn_Tr}),
\begin{align}
    &T(-x+d,y) = e^{-i\pmb{q}_1 \cdot \pmb{r}} \sigma_x T_1 \sigma_x + e^{-i\pmb{q}_2 \cdot \pmb{r}} e^{i(2\phi+q_{\text{\tiny{2x}}}d)} \sigma_x T_2 \sigma_x + e^{-i\pmb{q}_3 \cdot \pmb{r}} e^{e^{i(\phi-q_{\text{\tiny{2x}}}d)}} \sigma_x T_3 \sigma_x \\
    &\sigma_x T(x,y) \sigma_x = e^{-i\pmb{q}_1 \cdot \pmb{r}} \sigma_x T_1 \sigma_x + e^{-i\pmb{q}_2 \cdot \pmb{r}} \sigma_x T_2 \sigma_x + e^{-i\pmb{q}_3 \cdot \pmb{r}} \sigma_x T_3 \sigma_x
\end{align}
and we get 
\begin{equation}\label{tuneling_sigmax}
    \sigma_x T(x,y) \sigma_x = T(-x+d,y)
\end{equation}
where $d=a_{\text{\tiny{M}}}/ \sqrt{3}$ modulo $\sqrt{3}a_{\text{\tiny{M}}}$.  
In other words sublattice-exchange of the tunneling Hamiltonian is 
equivalent to reflection by the $y$-axis combined with a translation. 
Combining Eq.~(\ref{tuneling_sigmax_h0}) and (\ref{tuneling_sigmax}), we obtain
\begin{equation}
    \tau_z \sigma_x \mathcal{H}(x,y) \sigma_x \tau_z = -\mathcal{H}(-x+d,y)
\end{equation}
Similarly,
\begin{align}
    &T^\dagger(x,y) = e^{i\pmb{q}_1 \cdot \pmb{r}} T_1 + e^{i\pmb{q}_2 \cdot \pmb{r}} e^{-i\phi} T_2 + e^{i\pmb{q}_3 \cdot \pmb{r}} e^{i\phi} T_3 \\
    &T^*(-x+d,y) = e^{i\pmb{q}_1 \cdot \pmb{r}} T_1 + e^{i\pmb{q}_2 \cdot \pmb{r}} e^{-iq_{\text{\tiny{2x}}}d} T_2 + e^{i\pmb{q}_3 \cdot \pmb{r}} e^{iq_{\text{\tiny{2x}}}d} T_3 \\
    & T^\dagger(x,y) = T^*(-x+d,y)
\end{align}
$d=a_{\text{\tiny{M}}}/ \sqrt{3}$. If $m_1 = m_2$, then 
\begin{equation}
    \tau_x \mathcal{H}(x,y)\tau_x = \mathcal{H}^*(-x+d,y)
\end{equation}

Applying Bloch's theorem to Eq.~(\ref{eq_s1},\ref{eq_s2}), we see that 
$H(\pmb{k})$ satisfies
\begin{align}
    \label{sym1} &\tau_z \sigma_x H(k_x, k_y) \sigma_x \tau_z = - H(-k_x, k_y) \\ 
    \label{sym2} &\tau_x H(k_x, k_y) \tau_x = H^*(k_x, -k_y) 
\end{align}
As a result of Eq.~(\ref{sym1}), the eigenvalues and eigenvectors satisfy $\varepsilon_{\text{c}_i}(k_x, k_y) = -\varepsilon_{\text{v}_i}(-k_x, k_y)$, $\psi_{\text{c}_i}(k_x, k_y) = \tau_z \sigma_x \psi_{\text{v}_i}(-k_x, k_y)$. Here 
$i=1,2,...$ labels the $i$-th conduction (c$_i$) or valence (v$_i$) band counting from charge neutrality. For Eq.~(\ref{sym2}), $\varepsilon_{n}(k_x, k_y) = \varepsilon_{n}(k_x, -k_y)$, $\psi_{n}(k_x, k_y) = \tau_x \psi^*_{n}(k_x, -k_y)$, where $n$ label bands. 

Now let us define the orbital magnetization contribution due to mixing between 
bands $n$ and $n^\prime$:
\begin{equation}
    \mathfrak{M}^{nn^\prime} = \frac{e}{\hbar} \text{Im} \int_{\text{\tiny{MBZ}}} \frac{d^2\pmb{k}}{(2\pi)^2} \frac{\langle n|\partial_x H|n^\prime \rangle \langle n^\prime|\partial_y H|n \rangle}{(\varepsilon_n - \varepsilon_{n^\prime})^2} (\varepsilon_n + \varepsilon_n^\prime)
\end{equation}
When the chemical potential at neutrality is in the middle of the gap, and the Hamiltonian has been truncated to a finite number ($2N$)
of bands via a plane-wave expansion cut-off, the total orbital magnetization is
\begin{equation}\label{tot_M}
\begin{aligned}
    \sum\limits_{i=1}^N M^{\text{v}_i} 
    &= \sum\limits_{i,j=1, i\neq j}^N \mathfrak{M}^{\text{v}_i \text{v}_j} + \sum\limits_{i,j=1}^N \mathfrak{M}^{\text{v}_i \text{c}_j} \\
    &= \sum\limits_{i,j=1, i\neq j}^N \mathfrak{M}^{\text{v}_i \text{v}_j} + \sum\limits_{i=1}^N \mathfrak{M}^{\text{v}_i \text{c}_i} + \sum\limits_{i,j=1, i\neq j}^N \mathfrak{M}^{\text{v}_i \text{c}_j}
\end{aligned}
\end{equation}
The first term in the last expression of Eq.~(\ref{tot_M}) is zero because $\mathfrak{M}^{\text{v}_j \text{v}_i} + \mathfrak{M}^{\text{v}_i \text{v}_j} = 0$. The second term in the last expression of Eq.~(\ref{tot_M}) is also zero because $\varepsilon_{\text{c}_i} + \varepsilon_{\text{v}_i}$ is antisymmetric and $\text{Im} \langle \psi_{\text{v}_i} |\partial_x H| \psi_{\text{c}_i} \rangle \langle \psi_{\text{c}_i} | \partial_y H | \psi_{\text{v}_i} \rangle$ is symmetric when $k_x$ is reflected to $-k_x$. Similarly, we can also prove that $\mathfrak{M}^{\text{v}_j \text{c}_i} + \mathfrak{M}^{\text{v}_i \text{c}_j} = 0$. It follows that the total magnetization at mid-gap vanishes.

\section{S2. Heisenberg model estimate of spin magnetization at finite temperatures}
In mean-field theory, the spin magnetization of the Chern insulator state in MATBG 
is one Bohr magneton per moir\'e unit cell.  
We estimate thermal fluctuation corrections to the spin magnetization by starting from a square lattice 
ferromagnetic Heisenberg model
with Hamiltonian 
\begin{equation}
    H = -J\sum\limits_{\langle i,j \rangle} \pmb{S}_i \cdot \pmb{S}_j
    = -J\sum\limits_{\langle i,j \rangle} \Big[ \frac{1}{2}(S^+_iS^-_j + S^-_iS^+_j)+S^z_iS^z_j \Big].
\end{equation}
Here $S^{\pm}_i = S^x_i \pm S^y_i$ are spin raising and lowering operators, $\langle i,j \rangle$ labels a
nearest-neighbor bond, and the 
effective Heisenberg coupling can be extracted by comparing with microscopic theoretical estimates of magnon energies\cite{Fengcheng_tBLG_magnon, Kwan_tBLG_magnon, Ajesh_collective}.
There is no quantum fluctuation correction to the ground state spin-magnetization of the Heisenberg model,
but thermal fluctuations are important at finite temperature.  Indeed corrections to the Heisenberg model that 
break spin-rotational invariance are necessary for a finite magnetization to survive at finite temperatures 
in the two-dimensional systems of interest.

Magnetic anisotropy induced by spin-orbit coupling (SOC) or external magnetic fields 
limits the importance of thermal fluctuations for the spin-magnetization.
The spin Hamiltonians that add easy-axis single-ion anisotropy $H_D$, anisotropic exchange $H_\lambda$ 
and perpendicular external magnetic field $H_B$ contributions are respectively,
\begin{equation}\label{Eq_hamil}
\begin{split}
    H_D &= -J\sum\limits_{\langle i,j \rangle} \pmb{S}_i \cdot \pmb{S}_j - D \sum\limits_i (S^z_i)^2 \\
    H_\lambda &=
    -J\sum\limits_{\langle i,j \rangle} \pmb{S}_i \cdot \pmb{S}_j - \lambda \sum\limits_{\langle i,j \rangle} S^z_i S^z_j \\
    H_B &= -J\sum\limits_{\langle i,j \rangle} \pmb{S}_i \cdot \pmb{S}_j - \mu_{\text{B}} B \sum\limits_i S^z_i. 
   \end{split}
\end{equation}
In the external magnetic field case we have assumed that $\pmb{B}=-B\hat{z}$, and we have dropped the g-factor since 
we will replace $B$ by an effective magnetic field due to SOC below.

We can determine whether or not a large spin-polarization is maintained by applying a linearized spin wave approximation\cite{HP}
to the Heisenberg model.  The Heisenberg Hamiltonian then reduces to a model of quantized bosonic spin wave (magnon) excitations:
\begin{equation}
    H = \varepsilon_{\sub{0}} + \sum\limits_{\pmb{k}} \omega_{\pmb{k}} a^\dagger_{\pmb{k}} a_{\pmb{k}}
\end{equation}
where $a^\dagger_{\pmb{k}}(a_{\pmb{k}})$ are Holstein-Primakoff\cite{HP} bosons
creation (annihilation) operators and $\varepsilon_{\sub{0}}$ is the ferromagnetic ground state energy.
The magnon spectra corresponding to the Hamiltonians in Eq.(\ref{Eq_hamil}) are
\begin{equation}
\begin{split}
    \omega_{\pmb{k}}^D &= 2JS\sum\limits_{\pmb{\delta}} \big(1-\cos(\pmb{k}\cdot \pmb{\delta})\big) + DS \\
    \omega_{\pmb{k}}^\lambda &= 2JS\sum\limits_{\pmb{\delta}} \big(1-\cos(\pmb{k}\cdot \pmb{\delta})\big) + 2\lambda Sz \\
    \omega_{\pmb{k}}^B &= 2JS\sum\limits_{\pmb{\delta}} \big(1-\cos(\pmb{k}\cdot \pmb{\delta})\big) + \mu_{\text{B}}B
\end{split}
\end{equation}
where $\pmb{\delta}$ is nearest-neighbor lattice vectors and $z$ is the number of nearest neighbors.

Since there is no established mechanism for single-ion anisotropy or anisotropic exchange in
graphene, we focus on the case in which the magnon gap is created by a magnetic field.
As we explain below, an effective magnetic field is generated by spin-orbit interactions and orbital order
-- so $B$ is non-zero even if no magnetic field is applied.  
At low temperatures we can replace the magnon energy by its $k \rightarrow 0$ limit,
\begin{equation}\label{omega}
    \omega_{\pmb{k}}
    \rightarrow 2JSa^2k^2 + \mu_{\sub{B}}B
\end{equation}
The thermal average of magnon occupation number $n_{\pmb{k}}=a^\dagger_{\pmb{k}} a_{\pmb{k}}$ follows the Bose-Einstein distribution
\begin{equation}
    \langle n_{\pmb{k}} \rangle = \frac{1}{e^{\omega_{\pmb{k}/k_{\text{B}}T}}-1}
\end{equation}
Since the spin-magnetization is reduced by $1$ for each excited magnon in the linear spin wave approximation,
the spontaneous spin magnetization (per unit cell) $M$ as a function of temperature is
\begin{equation}\label{Eq_M}
    M(T) = S - \Delta M(T) = S - \frac{1}{N} \sum\limits_{\pmb{k}} \langle n_{\pmb{k}} \rangle,
\end{equation}
where $S=1/2$ is the ground state magnetization and $N$ is the number of lattice sites. 
$\Delta M(T)$ is the spontaneous magnetization contribution of magnons as a result of thermal fluctuations and can be rewritten in the energy integration
\begin{equation}\label{Delta_M1}
    \Delta M(T) = a^2 \int\limits_{\mu_{\text{B}}B}^{\varepsilon_{\text{zb}}} d\varepsilon \frac{\mathcal{D}(\varepsilon)}{e^{\varepsilon/k_{\text{B}}T}-1} 
    \end{equation}
where $a$ is the lattice constant, $\mathcal{D}(\varepsilon)$ is the magnon density-of-state 
per unit area, and $\varepsilon_{\sub{zb}}$ is the maximum magnon energy. 

When the effective magnetic field vanishes ($\mu_{\sub{B}}B=0$), the integral for $\Delta M$ in Eq.~(\ref{Delta_M1}) diverges logarithmically at any finite temperature, {\it i.e.} spins are not ordered. This property is consistent with 
the Mermin-Wagner theorem\cite{MerminWagner} which states that there is no spontaneous continuous symmetry breaking in a system with short-range interactions at any finite temperature for dimensionality $d \leq 2$.

\begin{figure*}[!h]
\centering
  \includegraphics[scale=0.8]{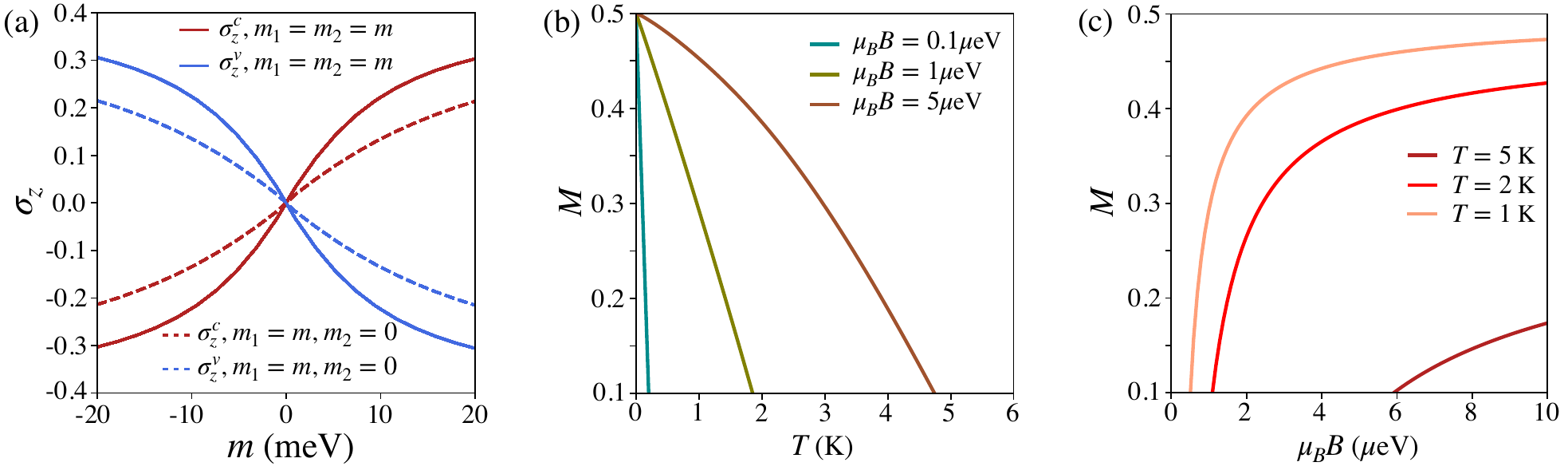}
  \vspace{-5pt}
  \caption{\label{figS1} {\small
  (a) Sublattice polarization $\sigma_z$ of $1.1^\circ$-TBG. When both graphene sheets are aligned with hBN in the same relative orientation ($m_1=m_2=m$, solid lines), the sublattice polarization is larger than when only one graphene sheet is aligned with hBN ($m_1=m$, $m_2=0$, dashed lines). Sublattice polarization of flat conduction band ($\sigma_z^\text{c}$, red) is always opposite with that of flat valence band ($\sigma_z^\text{v}$, blue).
  (b) $M$ as a function of temperature $T$ for three different effective magnetic fields $\mu_{\sub{B}}B=0.1,1$ and $5$ $\mu$eV.
  (c)$M$ as a function of $\mu_{\sub{B}}B$ for $T=1,2$ and $5$ K.
  }
  }
\end{figure*}

In graphene, an effective magnetic field acting on spins is induced by spin-orbit interactions when the system is valley polarized.
The effective magnetic field can be calculated by averaging the SOC term in the graphene 
Hamiltonian $H_{\text{so}} = \lambda_{\text{so}} \sigma_z \tau_z s_z$\cite{SOC_KaneMele, SOC_G_Min, G_SOC_Yao}
over orbital states in the spin-polarized band: $\mu_{\sub{B}} B  = \lambda_{\text{so}} \langle \sigma_z \tau_z \rangle $, where $\sigma_z$ is 
the pseudospin operator that measures sublattice polarization and $\tau_z = \pm 1$ is the pseudospin operator that 
measures valley polarization:  
\begin{equation}
    \sigma^{(n)}_z = \frac{A_m}{A} \sum\limits_{\pmb{k}} \big( P_{\sub{A}}^{(n)}(\pmb{k}) - P_{\sub{B}}^{(n)}(\pmb{k}) \big)
\end{equation}
where $A_m$ is the moir\'e unit cell area, $A$ is the sample area, and $P_{\sub{X}}^{(n)}(\pmb{k})$ is the expectation
value of projection onto sublattice $X$. Figure \ref{figS1}(a) plots the sublattice polarization of $1.1^\circ$-TBG as a function of the 
mass $m_l$ parameters used to account for the influence of the hBN substrate for both the 
flat conduction band ($\sigma_z^\text{c}$, red) and the flat valence band ($\sigma_z^\text{v}$, blue).
When both graphene sheets are aligned with hBN in the same relative orientation ($m_1=m_2=m$, solid lines), $\sigma_z^\text{c} \sim 0.2$ for a typical mass $m \sim 10$ meV. When only one graphene sheet is aligned with hBN ($m_1=m$, $m_2=0$, dashed lines), $\sigma_z^\text{c} \sim 0.1$ for $m \sim 10$ meV. Sublattice polarizations calculated in the  single-particle picture qualitatively agree with the results of self-consistent Hartree-Fock calculations which give $\sigma_z \sim 0.1$\cite{Ming_tBLG} at typical interaction strengths $\epsilon \sim 20$.

The effective SOC strength parameter for $\pi$ electrons in graphene has been  
estimated to be $\lambda_{\text{so}} \sim 1\mu$eV\cite{SOC_G_Min, G_SOC_Yao} using a tight-binding model with $s$ and $p$ orbitals. 
First principle calculations\cite{G_SOC_Konschuh, G_SOC_Boettger} that include $d$ orbitals estimate larger 
values $\lambda_{\text{so}} \in (25,50) \mu$eV.  Accounting for partial sublattice polarization and thermally suppressed valley polarization   
we can expect that $\mu_{\sub{B}}B = \lambda_{\text{so}} \langle \sigma_z \tau_z \rangle \in (0.1,5)$ $\mu$eV.
Spin-wave estimates of the thermal fluctuation correction to the spin magnetization 
are presented in Fig.~\ref{figS1}(b-c) where we plot $M$ calculated 
using the magnon dispersion in Ref.\cite{Ajesh_collective}.
Figure \ref{figS1}(b) shows $M$ as a function of temperature $T$ for 
$\mu_{\sub{B}}B=0.1, 1$ and 5 $\mu$eV.  The spin-wave estimate is accurate, of course, only when 
a substantial spin-magnetization is maintained.  We conclude from these calculations that the spin-magnetization 
is substantially reduced by thermal fluctuations, possibly to very small values depending on SOC strength. Figure \ref{figS1}(c) shows $M$ as a function of $\mu_{\sub{B}}B$ 
for $T=1,2$ and $5$ K. At $T = 2$ K, which is a typical temperature for MATBG QAHE transport measurements
(the anomalous Hall resistance remains accurately quantized up to $T \sim 3$ K in MATBG experiments\cite{Young2019QAHE}), 
$\mu_{\sub{B}}B$ would have to exceed $1.5$ $\mu$eV in order for a significant fraction of the spin-polarization
to survive thermal fluctuations.

In summary, spins are not ordered in MATBG at a few Kelvins because of extremely weak SOC.

\section{S3. Magneto-transport Characteristics and Electrical Magnetization Reversal}
Transport characteristics as a function of carrier density and magnetic field 
depend strongly on whether or not the electrical reversal, on which we have focused in this paper, is present.
In the following discussion we will identify magnetic states by their sense of valley polarization, distinguishing the 
two possibilities by referring to them as $+$ or $-$ states.

\begin{figure*}[!h]
\centering
  \includegraphics[scale=1]{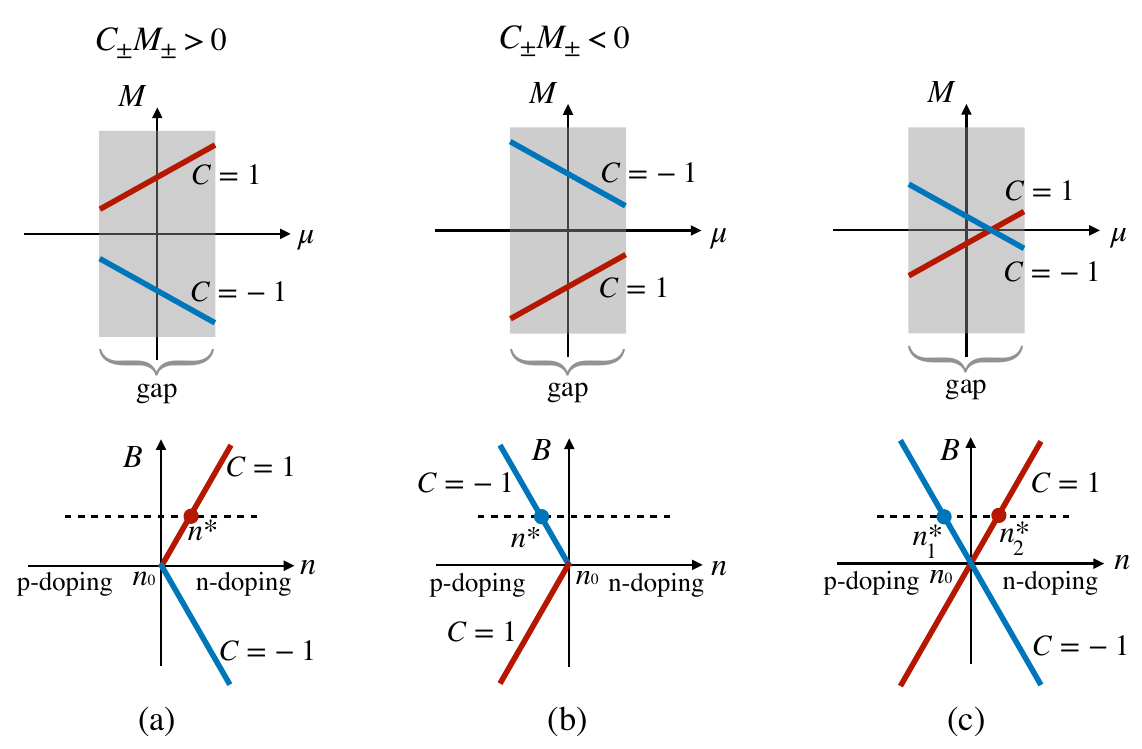}
  \vspace{-5pt}
  \caption{\label{figS2} {\small
  Chern insulator states can be identified by changes in the longitudinal resistivity at the density at which the gap opens.
  Using the St\v{r}eda formula, a Chern insulator can be identified with extension of the $\nu=C$ Shubnikov-de Haas Landau fan
  feature to $B=0$.  We note that positive magnetic fields favors states with positive magnetization. 
  (a) The sign of $M$ does not change across the band gap and is the same as the sign of Hall conductivity. The anomalous quantum Hall state is more robust for $n$-doping
  because the Chern gap moves into the conduction band when a magnetic field is applied.  
  (b) The sign of $M$ does not change across the band gap and is opposite to the sign of Hall conductivity. The anomalous quantum Hall state is then
  more robust for $p$-doping. (c) The sign of $M$ reverses across the band gap. The anomalous quantum Hall state is robust for both $n$- and $p$-doping,
  because the sign of the Chern number for a given magnetic field strength changes with carrier density.
  }
  }
\end{figure*}

Figure ~\ref{figS2} distinguishes three possible cases.  If $C_{+}M_{+} = C_{-}M_{-}  > 0$ for both signs of carrier density,
coupling to an external field will stabilize the positive Chern number state for positive field and the 
negative Chern number state for negative field.  On the other hand, If $C_{+}M_{+} = C_{-}M_{-} <0 $ a negative Chern number state
will be stabilized for positive fields and a positive Chern number state for negative fields.
It then follows from the St\v{r}eda formula\cite{Streda} for the magnetic field dependence of the density at which the
gap appears in the spectrum, $\partial n^*/\partial B = eC/h$, that the resistive anomaly associated with the gap will be stronger for one 
sign of carrier density.
As illustrated in the lower panels in
Fig.~\ref{figS2} the resistive anomaly is centered in the $n$-doped region for $C_{\pm}M_{\pm} > 0$, and 
in the $p$-doped region for $C_{\pm}M_{\pm} < 0$.  The longitudinal resistance anomaly associated with Chern insulator gaps is expected 
to extend over a finite region of carrier density around $n^*$, for example the region over which band-edge quasiparticles are localized.
The St\v{r}eda formula may be interpreted as saying that one of the Landau fan gaps visible in the Shubnikov-de Haas 
oscillations at stronger fields, the one at filling factor $\nu=C$, survives to zero magnetic field.  
As illustrated in Fig.~\ref{figS2}(c), the Landau fan structure 
is expected to extend to $B=0$ for both $n$- and $p$-doping when $C_{\pm}M_{\pm}$ changes sign across the gap.   

\begin{figure*}[!h]
\centering
  \includegraphics[scale=0.9]{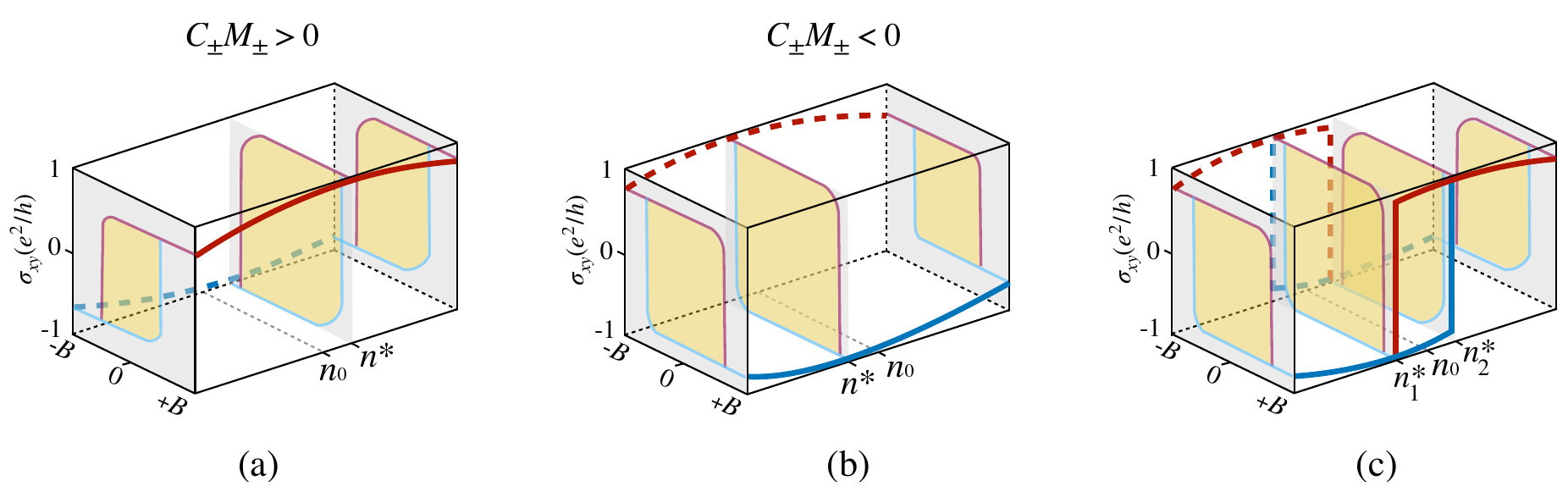}
  \vspace{-5pt}
  \caption{\label{figS3} {\small
  Schematic Hall conductivity hysteresis loops as sweeping the magnetic field at several carrier densities, and schematic Hall conductivity (solid and dashed lines) as a function of carrier density at magnetic fields on opposite extremes of the hysteresis loops.
  (a-b) Corresponding to Fig.~\ref{figS2}(a-b) respectively, the Hall conductivity does not change sign when the system is tuned from $p$-doping to $n$-doping for a fixed magnetic field. (c) Corresponding to Fig.~\ref{figS2}(c), the Hall conductivity changes sign when the system is tuned from $p$-doping to $n$-doping for a fixed magnetic field and is exactly quantized with opposite signs at two different densities $n_1^*$ and $n_2^*$. 
  }
  }
\end{figure*}

Changes in sign of $M_{\pm}C_{\pm}$ across the gap are also manifested in Hall resistivity measurements, as schematically
illustrated in Fig.~\ref{figS3} which plots
the Hall conductivity hysteresis loops as sweeping the magnetic field at several carrier densities. $n_0$ is the carrier density at which a non-trivial gap opens at zero magnetic field and $n^*$ is the shifted carrier density charactering the gap at finite magnetic field, as shown in lower panels of Fig.~\ref{figS2}. The Hall conductivities (solid and dashed lines) as a function of carrier density at the positive $B$ and negative $B$ extremes of the magnetic hysteresis loops are also shown in Fig.~\ref{figS3}, where quantized $\sigma_{xy}$ is achieved at $n^*$.
If the magnetization reverses sign across the band gap the quantized Hall conductivity
observed for a given sign of magnetic field also reverses (Fig.~\ref{figS3}(c)).

\end{document}